\documentclass[12pt]{JHEP}
\usepackage{amsmath,amssymb,amsfonts}
\usepackage[dvips]{graphicx}
\usepackage{epsfig}

\newcommand{\be}{\begin{equation}}
\newcommand{\ee}{\end{equation}}
\newcommand{\ba}{\begin{eqnarray}}
\newcommand{\ea}{\end{eqnarray}}
\newcommand{\bea}{\begin{array}}
\newcommand{\eea}{\end{array}}

\newcommand{\CM}{{\cal M}}
 
\newcommand{\CL}{{\cal L}} 
\newcommand{\CN}{{\cal N}}

\newcommand{\CP}{{\cal P}}
\newcommand{\Tr}{{\rm Tr}}

\newcommand{\bsigma}{{\boldsymbol{\sigma }}}
\newcommand{\bmu}{{\boldsymbol{\mu}}}
\newcommand{\bomega}{{\boldsymbol{\omega}}}
\newcommand{\bepsilon}{{\boldsymbol{\epsilon}}}
\newcommand{\bv}{{\bf v}}
\newcommand{\bW}{{\bf W}}
\newcommand{\bw}{{\bf w}}
\newcommand{\br}{{\bf r}}
\newcommand{\bx}{{\bf x}}
\newcommand{\bR}{{\bf R}}
\newcommand{\bq}{{\bf q}}
\newcommand{\bt}{{\bf t}}

\makeatletter \@addtoreset{equation}{section} \makeatother

\preprint{KIAS-P06001,hep-th/0605214}

\title{\Large \bf New $AdS_4\times X_7$ Geometries \\
with $\CN=6$ in M Theory}

\author{Ki-Myeong Lee\\
\\
School of Physics, Korea Institute for Advanced Study, Seoul
130-012
KOREA\\
\email{klee@kias.re.kr} }

\author{Ho-Ung Yee \\
\\
School of Physics, Korea Institute for Advanced Study, Seoul
130-012
KOREA\\
\email{ho-ung.yee@kias.re.kr} }

 \abstract{We study
supersymmetric $AdS_4\times X_7$ solutions of 11-dim supergravity
where the tri-Sasakian space $X_7$ has generically $U(1)^2\times
SU(2)_R$ isometry. The compact and regular 7-dim spaces
$X_7=S(t_1,t_2,t_3)$ is originated from 8-dim hyperkahler quotient
of a 12-dim flat hyperkahler space by $U(1)$ and belongs to the
class of the Eschenburg space. We calculate the volume of $X_7$
and that of the supersymmetric five cycle via localization. From
this we discuss the 3-dim dual superconformal field theories with
$\CN=3$ supersymmetry.}

\begin{document}

\section{Introduction and Conclusion}

The AdS-CFT correspondence predicts that the type IIB-theory on
the supergravity solution $AdS_5\times S^5$ with appropriate
5-form field strength is dual to a $\CN=4$ supersymmetric 4-dim
$SU(N)$ gauge theory, which is a superconformal field
theory\cite{Maldacena:1997re}. When the space $S^5$ is replaced by
a 5-dim Sasaki Einstein space, the dual gauge theory has less
supersymmetry with more complicated group and matter
structure\cite{Kehagias:1998gn}. These conformal theories can be
regarded as a field theory on a stack of D3 branes sitting at the
the singular tip of a Ricci flat 6-dim cone whose base is the
Sasaki-Einstein
space\cite{Klebanov:1998hh,Acharya:1998db,Morrison:1998cs}.

The AdS-CFT correspondence for the Freund-Rubin form, $AdS_4\times
X_7$, of a supersymmetric solution of 11-dim supergravity implies
that the M-theory on such background is dual to a supersymmetric
3-dim superconformal field theory. Again the field theory arises
as the SCFT on a stack of M2 branes at the singular apex of 8-dim
Ricci-flat space with special holonomy, whose base is 7-dim $X_7$.
One has to have at least one Killing spinor on $X_7$ to have a
special holonomy. Recently there have been found several countable
series of Sasaki-Einstein space in 5-dim and 7-dim, and their
AdS-CFT correspondence has been
studied\cite{Gauntlett:2004zh,Gauntlett:2004yd,Lu:2004ya}.
Especially the corresponding 3-dim SCFT's have $\CN=1$, or $2$
supersymmetries.

In this work we focus on a class of countable series of the
$AdS_4\times X_7$ spaces, which has not been studied before. The
dual SCFT has $\CN=3$ and is the SCFT on the M2 branes on the
singular tip of the 8-dim hyperkahler cone with $Sp(2)$ holonomy
and with its base being a tri-Sasakian space $X_7$. The 8-dim
hyperkahler cone is obtained by a hyperkahler quotient of
$\mathbb{R}^{12}$ by a U(1) symmetry group\cite{Hitchin:1986ea}.
(In general, one could have started from flat $\mathbb{R}^{4n +8}$
space with $U(1)^n$ hyperkahler quotient with $n\ge 1$ but here we
restrict to $n=1$ case for simplicity.) Our tri-Sasakian space
$X_7(\bt)=S(t_1,t_2,t_3)$ is characterized by three natural
numbers $t_1,t_2,t_3$ and has $SU(2)_R\times U(1)^2$,
$SU(2)_R\times SU(2)\times U(1)$ or $SU(2)_R\times SU(3)$ isometry
depending on none, two,  or all of $t_a$ coincide, respectively.
We calculate the volume of the $X_7$ and its supersymmetric
5-cycles $\Sigma_5$ and obtain a rational expression for the ratio
of the volume of $X_7$ and that of the unit 7-sphere and so on. We
also discuss the dual 3-dim superconformal field theories with
$\CN=3$. During our investigation, we found these type of space
has appeared before in the mathematics
literature~\cite{Boyer:1998sf,bgm2,Bielawski} where it is known as
the Eschenburg space~\cite{Esch1}. However, our calculation of the
volumes of the space and the super 5-cycles  found in this paper
seems original.

The simplest case with $X_7=S(1,1,1)$ is known as $N(1,1)$ and its
cone is the relative moduli space of a single instanton in $SU(3)$
gauge group, which is hyperkahler 8-dim space with one scale and
the coset space $SU(3)/U(1)$. There has been considerable work on
the AdS-CFT correspondence on the $AdS_4\times N(1,1)$
space\cite{Fre':1999xp,Fabbri:1999hw,Billo:2000zr}. Especially as
$N(1,1)$ is homogeneous, one can study the Kaluza-Klein modes of
the theory to compare them with the dual SCFT. More recent
investigation on $AdS_4\times X_7$ space with known $X_7$ and its
marginal deformation  can be found in Ref.\cite{Gauntlett:2005jb}.

Our work has been motivated in part by the effort to understand
the mysterious 3-dim $\CN=16$ supersymmetric conformal field
theory which is the low energy theory of $N$ parallel M2 branes.
They can be regarded as the strong coupling limit $e^2\rightarrow
\infty$ of $\CN=16$ supersymmetric Yang-Mills theories, as one can
easily see in the M-theory limit of D2 branes. One may deform the
$\CN=16$ supersymmetric Yang-Mills by adding Chern-Simons terms,
so that the resulting theory has a less supersymmetry
$\CN=3$~\cite{Kao:1992ig}\cite{Kao:1993gs}. In the infrared limit
or strong coupling limit $e^2\rightarrow \infty$, the theory
becomes purely Chern-Simons Higgs, which is superconformal. While
the Chern-Simons level $k$ is quantized to be integer, the small
$|k|$ limit is the strong-coupling limit. Unfortunately its
physics is not well understood. One may still hope that the
physics near $k=0$ is similar to that of $N=16$ superconformal
theory. (See for a similar idea in Ref.\cite{Schwarz:2004yj}.)

Another motivation was to try to understand further the old result
on the superconformal field theory dual for the AdS geometry with
tri-Sasakian space $X_7=N(1,1)$ whose 8-dim cone is the relative
moduli space of a single instanton in $SU(3)$ gauge theory. While
the corresponding field theory may have some component of
Chern-Simons theory, the t'Hooft coupling from the geometry seems
to be related to the parameter $N$ of the corresponding gauge
group $SU(N)\times SU(N)$, instead of the generic 'tHooft coupling
$N/k$ of the Chern-Simons-Higgs theory where $k$ is the integer
quantized Chern-Simons level. Our models would provide more
examples along this line.

Final motivation was to try to construct new tri-Sasakian geometry
similar to instanton moduli space by generalizing the moduli space
of three distinct magnetic monopoles, which could be constituent
of a single instanton of $SU(3)$ theory in $R^3\times S^1$
geometry\cite{Lee:1997vp}. Thus one wants to generalize the moduli
space of $N$ distinct magnetic monopoles in $SU(N)$ theory broken
to $U(1)^{N-1}$ in $R^3\times S^1$. There would be magnetic
monopoles for each simple root of the extended Dynkin diagram of
$SU(N)$ gauge group. But we generalize the interaction strength
between nearest neighbor by arbitrary magnitude. Only requirement
is that the geometry is smooth whenever magnetic monopoles are
coming together except when all of them are coming together. In
appendix  this is shown to lead to the interaction strength
between each link to be some natural number instead of the unity
as in the $SU(N)$ case. In the limit where $N-1$ monopoles become
massless, the relative geometry has only one scale parameter which
controls the overall size of the system, and so a cone-like
geometry with a singularity at the apex of the cone. Our geometry
$X_7$  can also obtain from this approach.

The geometry we are interested in is the $AdS_4\times X_7$ type
solutions of $D=11$ supergravity where $X_7$ is an Einstein
manifold and the four-form field strength $F_4\sim vol_{AdS_4}$.
We normalize the metric on $X_7$ so that
$R_{\mu\nu}(X_7)=6g_{\mu\nu}(X_7)$. To preserve some of 32
supersymmetry of 11-dim supergravity, the eight dimensional cone
$\CM_8$ over $X_7$ with the metric
\be ds^2_{\CM_8} = dr^2 + r^2 ds^2(X_7) \ee
should be Ricci-flat and have special holonomy. (For example see
Ref.~\cite{Figueroa-O'Farrill:1999va}.) When the cone has
$Sp(2)=SO(5)$ holonomy, and so hyper-K\"ahler, the space $X_7$ is
tri-Sasakian and the dual theory is a $\CN=3$ SCFT.

Such a geometry arises as the near horizon limit of $M2$ branes
lying at the singular apex of the Ricci-flat cone $\CM_8$. The
dual SCFT lives on the $M2$ branes. The flux of $F_4$ on $X_7$ is
proportional to the number of $M2$ branes. Baryonic states of SCFT
are dual to five-branes wrapping five-cycles $\Sigma_5$ in the
manifold $X_7$. For supersymmetirc states the five-cycles must
lift to supersymmetric 6-cycles in the cone $\CM_8$. A
supersymmetric 6-cycle will be holomorphic with respect to one of
the three complex structures, breaking two of six supersymmetries.
The dimension of the baryonic operators are given by the geometric
formula~\cite{Fabbri:1999hw}
\be \Delta = \frac{\pi N}{6}\frac{{\rm Vol}(\Sigma_5)}{{\rm
Vol}(X_7)} \, . \label{baryonn}\ee
In SCFT, they can often be predicted from their R-charges.
Comparing the two predictions is then a non-trivial check for the
gauge/gravity correspondence.

This Eschenburg space $X_7=S(t_1,t_2,t_3)$~\cite{Boyer:1998sf} can
be regarded as a left-quotient space of $SU(3)$ group manifold by
$U(1)$ group whose elements are ${\rm diag}(e^{it_1\psi},
e^{it_2\psi},e^{it_3\psi})$. Their homological properties seem to
be known. Here we provide an explicit metric and calculation of
the volume of the space and supersymmetric 5-cycles. The space
$X_7=S(t_1,t_2,t_3)$ depends only on the $t_a$'s up to overall
common factor. It is homogeneous with $SU(3)\times SU(2)_R$
symmetry when $t_1=t_2=t_3$. When $t_1=t_2\neq t_3$, the space has
co-homogeneity one with $SU(2)\times U(1)\times SU(2)_R$ symmetry.
When all $t_a$ are different from each other, the space has
co-homogeneity two with $U(1)^2\times SU(2)_R$ symmetry. We find
the two kind of expressions for the metric for $X_7(\bt)$. The
first one is explicit but not useful. The second one is  more
implicit but shows the  symmetric and cone structures clearly.

Instead of using the metric, we use the equivarient cohomology and
localization technique \cite{Moore:1997dj} to calculate the volume
$X_7(\bt)$. This approach is somewhat esoteric and so the detail
is provided here. We find that the ratio of the volume of the
tri-Sasakian space $X_7$ and that for any supersymmetric 5-dim
cycle $\Sigma_5(\bt)$ is independent of $\bt=(t_1,t_2,t_3)$ and
identical to the ratio for the volume of unit 7-sphere and that of
unit 5-sphere. The explicit form for the volumes are
\be
\frac{vol(X_7(\bt))}{vol(S_7)}=\frac{vol(\Sigma_5(\bt))}{vol(S_5)}
=\frac{t_1t_2t_3(t_1t_2+t_2t_3+t_3t_1)}{l.c.m.(t_1t_2,t_2t_3,t_3t_1)(t_1+t_2)(t_2+t_3)(t_3+t_1)}
\, , \ee
where $l.c.m.$ means the least common multiple and
$vol(S_7)=\pi^4/3$ for the unit 7 sphere and $vol(S_5)=\pi^3$ for
the unit 5 sphere. The maximum of the above ratio for any $\bt$
appears when $t_1=t_2=t_3=1$ for the well-known space $N(1,1)$.

The dual superconformal field theory in three dimension is
$SU(N)_1\times SU(N)_2$ gauge theory with $\CN=3$ supersymmetry.
The matter fields $U_a=(u_a, -v_a^*)$ are hypermultiplets in
$\CN=4$ language belonging to the symmetrized product
representation $Sym^{t_a}(N)$ of the fundamental representation of
$SU(N)_1$ and that $Sym^{t_a}(\bar{N})$ of the anti-fundamental
representation of $SU(N)_2$. The internal global symmetry is again
$SU(2)_R\times U(1)^2$ for distinct $t_a$. The chiral primary
operators and baryonic operators show that one can assign a chiral
dimension 1/2 for the $U_a$ field and its complex conjugates.

There are several directions to pursue. Our SCFT is again
mysterious as the $\CN=8$ superconformal field theory in 3-dim as
it is not quite the Chern-Simons theory.  They may be the
strong-coupling limit of the supersymmetric Yang-Mills
Chern-Simons theory with $\kappa \rightarrow 0$. One curious
aspect of our AdS-CFT correspondence  is that there is no obvious
geometry for the dual theory when $t_1=t_2=t_3\ne 1$ as the
$X_7(\bt)$ is defined only up to common factors of $t_a$.

The plan of the paper is as follows. In Sec.2 we define the 8-dim
hyperkahler space $\CM_8(\bt)$ which is a singular cone and is
obtained from a hyperkahler quotient of 3-dim quaternion space
$\mathbb{H}^3=\mathbb{R}^{12}$ by using a single $U(1)$ group. We
find its metric explicitly and also show the space has the cone
geometry. We identify its isometry. In Sec.3 we review the
homology property of tri-Sasakian space $X_7(\bt)$ first. Then we
calculate the volumes of $X_7(\bt)$ and supersymmetric 5-cycles
$\Sigma_5$ in $X_7$ in the language of equivariant cohomology. In
Sec.4, we identify the dual SCFT and study its properties. In
Appendix, we generalize the caloron moduli space.

\section{Hyperkahler Space $\CM_8(\bt)$}

Let us start from 12-dim flat hyperkahler space $\mathbb{H}^3$
defined by the three quaternions $q_1,q_2,q_3$. (See for example
Ref.~\cite{Gibbons:1996nt} for an introduction.) Each quaternion
is defined as
\be q_a = q^4_a+ i\bsigma\cdot \bq_a \, , \;
\bar{q}_a=q_a^4-i\bsigma\cdot\bq_a\, , \ee
with four real numbers $q^\mu_a, \mu=1,2,3,4$ and three Pauli
matrices $\sigma^1,\sigma^2,\sigma^3$. The Euclidean flat metric
on 12-dim is
\be ds^2 = \sum_a \frac{1}{2} {\rm tr}(dq_a\otimes d\bar{q}_a) =
\sum_a dq_a^\mu dq_a^\mu \, , \ee
and the three K\"ahler forms are
\be \bomega\cdot \bsigma = \frac{1}{2} dq \wedge d\bar{q} \, .
\label{kahlerforms}\ee
Sometimes we use complex coordinates for quaternions as
\be q_a = \left( \begin{array}{rr} u_a & v_a \\ -\bar{v}_a &
\bar{u}_a \end{array}\right)\, , \label{uvcoord}\ee
in which the metric becomes
\be ds^2 =\frac{1}{2}\sum_a (du_a \otimes d\bar{u}_a+ dv_a \otimes
d\bar{v}_a +c.c.)\, .\ee
Another useful coordinate for quaternions is
\be q_a = p_a e^{i\sigma_3 \psi_a} \, , \ee
where $p_a$ is pure imaginary, or $\bar{p}_a=-p_a$. In terms of
the 3-dim Cartesian coordinates $\br_a$ such that
\be i{\bf r}_a\cdot \bsigma = q i\sigma_3 \bar{q} =-ip_a \sigma^3
p_a\ee
and angle variable $\psi_a$, the flat metric on 12-dim becomes
\be ds^2 = \frac{1}{4}\sum_a \left(\frac{d{\bf r}_a^2}{r_a} + r_a
(d\psi_a +{\bf w}_a \cdot d{\bf r}_a)^2 \right)\, , \label{flat}
\ee
where $r_a=|{\bf r}_a|$ and $\nabla\times {\bf w}_a({\bf r}_a)=
\nabla(1/r_a)$.

For each triple natural numbers $t_1,t_2,t_3$, we consider a
corresponding abelian symmetry $U_\bt(1)$, under which
\be q_a \rightarrow q_ae^{i\sigma_3 t_a \chi }\, ,\quad a=1,2,3 \,
\, .\ee
The $U_\bt(1)$ is unique up to a common factor on triples. The
corresponding moment map $\bmu$ is
\ba \bmu \cdot \bsigma &=& \sum_a t_a q_a \sigma_3 \bar{q}_a
= \sum_a t_a {\bf r}_a \nonumber\\
&=& \sum_a t_a \left( \begin{array}{cc} |u_a|^2-|v_a|^2 & -2u_av_a \\
- 2\bar{u}_a \bar{v}_a & -|u_a|^2+|v_a|^2 \end{array}\right)\, .
\label{momentmap} \ea
The space which satisfies the constraint $\bmu=0$ becomes
9-dimensional. Once we mod out $U(1)_t$ on this space, the
resulting quotient space becomes 8-dim hyperkahler space,
\be {\cal M}_8(\bt) = \bmu^{-1}(0)/U_\bt(1) \, .\ee
This process of hyperkahler quotient is defined by three natural
numbers $t_1,t_2,t_3$. As ${\cal M}_8(\bt)$ is hyperkahler, it is
Ricci-flat automatically.

Let us consider in detail the symmetry of the hyperkahler space
$\CM_8(\bt)$. The first one is the $SU(2)_R$ symmetry which
rotates three complex structures in 12-dim space,
\be q_a \rightarrow \exp(-\frac{i}{2}\bepsilon\cdot\bsigma) q_a,
\; a=1,2,3 \, , \ee
where $\bepsilon$ are the $SU(2)$ parameters. Under this $SU(2)$
transformation, ${\bf r}_a$ for each $a$ transforms as a vector.
This is commuting with the hyperK\"ahler quotient and so the
resulting space has $SU(2)_R$ symmetry. The additional symmetry
arises from the transformation
\be q_a \rightarrow q_b \bigl(\exp{iT \sigma^3} \bigr)_{ba} \ee
where $T$ is the $U(3)$ generator which commutes with the
$U(1)_\bt$ generator
\be {\bf t}= {\rm diag}(t_1,t_2,t_3) \, ,  \ee
and leaves $U_\bt(1)$ invariant subspace invariant. Thus when
$t_1=t_2=t_3$, the resulting symmetry is $SU(3)$. When
$t_1=t_2\neq t_3$, the resulting symmetry is $SU(2)\times U(1)$.
When $t_1,t_2,t_3$ are all different, the resulting symmetry would
be $U(1)^2$.

For $t_1=t_2=t_3$, the resulting 8-dim space is the moduli space
of a single SU(3) instanton in the center of mass frame. The 8
parameters denote a single scale parameter and 7 coordinates for
the coset space $SU(3)/U(1)$, and so the space is cone-like. For
generic $t_a$, the metric is more complicated. A simple way to
write the metric is to start from the flat metric (\ref{flat}) and
express the ${\bf r}_3$ in terms of ${\bf r}_1$ and ${\bf r}_2$ by
using the moment map (\ref{momentmap}) so that with $A=1,2$
\be 4ds^2 = C_{AB} d{\bf r}_A d{\bf r}_B + C^{AB}(d\psi_A + {\bf
w}_{AC}\cdot d{\bf r}_C)(d\psi_B+{\bf w}_{BD}\cdot d{\bf r}_D)\, ,
\ee
where
\ba && C_{11}=\frac{1}{r_1}+\frac{t_1^2}{t_3|t_1{\bf r}_1+t_2 {\bf
r}_2|}\nonumber \, ,
\\
&& C_{22}= \frac{1}{r_2}+ \frac{t_2^2}{t_3|t_1{\bf r}_1+ t_2 {\bf
r}_2|} \nonumber \, ,\\
&& C_{12}=C_{21} =\frac{t_1t_2}{t_3|t_1{\bf r}_1+t_2{\bf r_2}|}\,
,\ea
and the vector potential satisfies $\nabla_C \times {\bf w}_{AB}=
\nabla_C C_{AB}$. This metric is hyperkahler and regular unless
$r_1=r_2 =0$ simultaneously\cite{pedersen,Lee:1996kz}.

To express the metric so that the cone-structure is manifest needs
more work. Let us focus on the generic case where all $t_a$ are
different. The moment map vanishes and so $t_1{\bf r}_1 +t_2{\bf
r_2}+t_3{\bf r}_3=0$, which defines a triangle whose side length
are  $t_1r_1, t_2r_2,t_3r_3$. Using the $SU(2)_R$ transformation,
we can put this triangle on the 1-2 plane. In this case, the
complex coordinates $u_a$ and $v_a$ satisfy
\be |u_a|= |v_a|=\sqrt{\frac{r_a}{2}}\, ,  \;\; \sum_{a=1}^3 t_a
u_a v_a = 0 \, . \label{mapcond}\ee
The general configuration would be made of the rotation of the
triangle in space and also the phase rotation of $u_a$ and $v_a$
in opposite way. Triangle has three independent parameters. The
spatial rotation has three independent parameters. The relative
phase of $u_a, v_a$ variables has three independent parameters,
one of which is the global $U(1)$ which should be mode out. Thus
there are eight independent parameters.

To specify the moduli parameter for the triangle, we choose the
parameters to be $u_a =v_a=\sqrt{r_a/2}= |u_a|e^{i\varphi_a/2}$,
which implies the moduli metric of the triangle to be
\be ds^2_{\Delta} = \frac{1}{4}\sum_{a=1}^3\biggl( \frac{1}{r_a}
dr_a^2 + r_a d\varphi_a^2 \biggr)\, ,\ee
with the condition (\ref{mapcond}) being
\be \sum_a t_a r_a e^{i\varphi_a}=0 \, .  \label{triangle} \ee
This is the condition on for three complex vectors $t_ar_a
e^{i\varphi_a}$ to form a triangle. This constraint depends only
on the relative angles $\theta_a$ of vectors as
\be \theta_1 = \varphi_3-\varphi_2 \, , \;\; \theta_2=
2\pi+\varphi_1-\varphi_3\, , \;\;   \theta_3 =
\varphi_2-\varphi_1\, ,  \ee
where $0\le \varphi_a<2\pi$. Only two of the relative angles are
independent as $\theta_1+\theta_2+\theta_3=2\pi$. The overall
orientation angle $\varphi =
\frac{\varphi_1+\varphi_2+\varphi_3}{3}$ of the triangle are a
part of the rotational degrees from $SU(2)_R$. The above triangle
condition (\ref{triangle}) implies the three following conditions
on the length and relative angles as given in elementary geometry:
\ba && t_1^2r_1^2= t_2^2r_2^2+t_3^2r_3^2+2t_2r_2t_3r_3
\cos\theta_1
\nonumber \; ,\\
&& t_2^2r_2^2= t_3^2r_3^2+t_1^2r_1^2+2t_3r_3t_1r_1\cos\theta_2
\nonumber
\;,\\
&& t_3r_3^2= t_1^2r_1^2+t_2^2r_2^2+2t_1r_1t_2r_2\cos\theta_3 \; ,
\label{triangle1} \ea
of which only two are independent. Thus these conditions reduce
the independent variables to three, which we choose as one length
variable, and two relative angle variables.

To solve the above constraints (\ref{triangle1}), let us introduce
an angle variable $A$, a length square variable $L$, and an area
variable $S$ such that
\ba &&
A= -(\cot\theta_1 +\cot\theta_2+\cot\theta_3)\; , \nonumber \\
&& L= \sum_a t_a^2 r_a^2\; , \nonumber \\
&& S= t_1r_1 t_2r_2
\sin\theta_3=t_3r_3t_1r_1\sin\theta_2=t_2r_2t_3r_3\sin\theta_1 \;
.\ea
Note that $S$ is twice the area of the triangle, and the triangle
condition implies that
\be S=\frac{L}{2A}\quad ,\;\; \;\; t_a^2 r_a^2 = L \biggl(
1+\frac{\cot \theta_a}{A} \biggr)\;\;\;\; {\rm for \;\; each }\;\;
a \ee
with $A\ge 0$. Let us now introduce the radial variable in 12-dim
flat space as the length variable,
\be r = \sqrt{ |q_1|^2+|q_2|^2+|q_3|^2} =\sqrt{r_1+r_2+r_3}\, .\ee
Defining a function $B$ of angles $\theta_a$ as
\be B = \sum_a \frac{1}{t_a} \sqrt{\biggl( 1-
\frac{\cot\theta_a}{\cot\theta_1 +\cot\theta_2+\cot\theta_3}
\biggr)} \, ,\ee
we see the length variable $L$ is given in terms of three
independent variables $r,\theta_a$ as follows,
\be L = \frac{r^4}{B^2} \, .\ee
So the variables $r_a$ can be written in terms of scale variable
$r$ and angles $\theta_a$ as follows,
\ba && r_a = r^2 \rho_a \, ,\;\; \rho_a \equiv \frac{1}{t_a}
\sqrt{\biggl(1-\frac{\cot\theta_a}{\sum_b \cot\theta_b} \biggr)}
\biggl/ \biggr. \sum_c
\frac{1}{t_c}\sqrt{\biggl(1-\frac{\cot\theta_c}{\sum_d
\cot\theta_d} \biggr)}\, . \ea
Note that three functions $\rho_a$ of angles $\theta_a$ satisfies
the condition $\rho_1+\rho_2+\rho_3=1$.

The moduli space of the triangle on the plane would be then
\be ds^2_{\Delta}=\frac{1}{4} \sum_a \left(\frac{1}{r_a}dr_a^2 +
r_a d\varphi_a^2 \right)= dr^2 + \frac{r^2}{4} \sum_a \biggl(
\frac{d\rho_a^2}{\rho_a} + d\varphi_a^2\biggr)\, ,
\label{metrictri} \ee
as $\sum \rho_a = 1$. Note the $\varphi_a$ can be written as the
relative angles $\theta_a$ and the overall orientation of the
triangle on the plane. The metric for the $\CM_8$ can be now
obtained by parameterizing quaternion as follows
\be Q\equiv \biggl( \begin{array}{ccc} u_1 & u_2 & u_3 \\
-\!\bar{v}_1 & -\bar{v}_2 & -\!\bar{v}_3 \end{array} \biggr) =
RQ_0 T\, ,  \ee
where the $R$ is an $SU(2)$ element parameterized by Euler angle,
which includes the orientation of the triangle on the plane, $T$
is a diagonal $U(3)$ element, say, $T= {\rm
diag}(e^{i\psi_1},e^{i\psi_2}, e^{i\psi_3})$ and $Q_0$ is the
value of $Q$ when the triangle is on 1-2 plane and so
\be Q_0 = \biggl(
\begin{array}{ccc} \sqrt{r_1}e^{i\varphi_1/2} &
\sqrt{r_2}e^{i\varphi_2/2} &
\sqrt{r_3}e^{i\varphi_3/2} \\
-\sqrt{r_1}e^{-i\varphi_1/2} &- \sqrt{r_2}e^{-i\varphi_2/2} & -
\sqrt{r_3}e^{-i\varphi_3/2} \end{array}\biggr) \, .\ee
The metric of the triangle is $ds^2_{\Delta} = dQ_0 d\bar{Q}_0$ of
the metric (\ref{metrictri}) and so the metric on the 9-dim space
is
\be ds^2_{\bmu^{-1}(0)} = \frac{1}{2} ( dQ\otimes d\bar{Q}+
d\bar{Q}\otimes dQ )\; . \label{metric9} \ee
It is trivial to mode out $U_\bt(1)$ to get the 8-dim hyperkahler
space $\CM_8(\bt)$. The 3 kahler forms are again given by
\be \bomega\cdot\bsigma = dQ \wedge d\bar{Q}\, . \ee

The isometry of $\CM_8(\bt)$ can be easily read. First of all the
$SU(2)$ transformation by $R$ matrix leads to $SU(2)_R$ symmetry
which mixes three complex structure. In addition there are
$U(1)\times U(1)$ isometries from the tranformations given by $T$
matrix modulo $U_\bt(1)$, which are tri-holomorphic as they leave
three kahler structures invariant. When some of $t_1, t_2,t_3$
become identical, these tri-holomoprhic isometries get enhanced.
If only two of three are identical, $U(1)^2$ gets enhanced to
$U(1)\times SU(2)$. If all three $t_a$ are identical, $U(1)^2$
gets enhanced to $SU(3)$.

The metric (\ref{metric9}) has no mixed terms for $dr$ and other
angle variables including $dR$, $dT$ as $R$ and $T$ are unitary.
Thus, the metric on the hyperK\"ahler space $\CM_8$ has a cone
structure,
\be ds^2_{{\cal M}_8}= dr^2 + r^2 ds_{X_7}^2\; . \ee
It is also regular everywhere except at the tip of the cone. It is
Ricci-flat and the 7-dim space $X_7$ is smooth everywhere without
singularity and characterized by three natural numbers
$(t_1,t_2,t_3)$ without common factor. In the next section, we
study properties of this 7-dim tri-Sasakian space $X_7(\bt)$.

\section{Tri-Sasakian Space $X_7({\bf t})$}

We have now the 8-dimensional hyperkahler space $\CM_8(\bt)$,
whose metric is cone-like and determined by three natural numbers
$t_1,t_2,t_3$ modulo common factor. This space is smooth except at
the tip of the cone. As the metric is written down almost
explicitly, one has now the corresponding 7-dimensional
tri-Sasakian space $X_7({\bf t})$ which is smooth everywhere.

There are several properties of this space which are relevant for
our consideration. The isometry of $X_7(\bt)$ is identical to the
cone geometry $\CM_8(\bt)$. With $t_1=t_2=t_3$, the unique space
is equivalent to $S(1,1,1)=N(1,1)$. With $t_1=t_2\neq t_3$, there
are class of geometry with $S(r,r,s)$ with coprime natural numbers
$r,s$. Finally, when all $t_a$ are different, we can assume that
there is no common factor in them. This space
$X_7(\bt)=S(t_1,t_2,t_3)$ is non-singular. It is sometimes called
the Eschenburg space. This toric Sasakian space is a subfamily of
the more general spaces, bi-quotients of U(3) group manifold,
which was studied by Eschenburg\cite{Esch1}.

The quotient of $S(t_1,t_2,t_3)$ by $SO(3)_R$ action is a
quaternionic K\"ahler orbifold. For any $U(1)_R$ subgroup of
$SO(3)_R$, one can locally write the metric as
\be ds^2(X_7) = (d\psi+\sigma)^2 + ds^2(M_6)\, , \ee
where $M_6$ is locally Kahler-Einstein. If the Reeb vector
$\partial/\partial\psi$ has a closed orbit, then $M_6$ is in
general Kahler orbifold. A tri-Sasakian space is regular if its
quotient by $SO(3)_R$ is a quaternionic K\"ahler manifold. Our
case $S(t_1,t_2,t_3)$ would be a smooth tri-Sasakian 7-manifold
which is not regular unless $t_1=t_2=t_3$ in which case it is
homogeneous. Indeed all homogeneous tri-Sasakian spaces in $4n+3$
dimensions seem to be associated with Lie algebra and seem to be
originated from the moduli space of a single instanton in a gauge
theory of a given Lie group as it has only one scale parameter and
is hyperkahler. The betti-numbers of the space
$S(t_1,t_2,t_3)$~\cite{Boyer:1998sf} are
\be b_0=b_7=1\, , b_1=b_6=0\, , b_2=b_5=1, b_3=b_4=0\, ,  \ee
which indicates one can have nontrivial wrapping of the geometry
by $M2$ branes and M5 branes. The wrapping of 5 cycles by M5
branes leads to the baryonic objects in the dual SCFT.

One can obtain more general 7-dimensional toric tri-Sasakian space
first proposed in Ref.\cite{bgm2} by considering $N+2$-quaternion
space $q_a, a=1,2,...,N+2$ with $N$ independent $U(1)_t$ groups
acting on them with charge matrix
\be q_a \rightarrow q_a e^{i\sigma_3 t^A_a\chi_A}\, ,\ee
where $A=1,2,...N$ leads to $N$ abelian symmetry. There are $N$
corresponding moment map
\be \mu^A = \sum_{a=1}^{N+2} t_a^A q_a \sigma_3 q_a\, . \ee
The hyperKahler quotient of the $N+2$ dimensional quaternion space
by these abelian groups leads to 8-dimensional hyperkahler space.
Since we propose dual SCFT only for $N=1$ case in this work,
generalization for $N\ge 2$ would be an interesting future
problem.

\subsection{The volume of tri-Sasakian space $ X_7(\bt)$ }

{\it The Volume of hyperkahler quotient $\CM$ and their
Equivariant Deformation}

Suppose that a hyperk\"ahler manifold $\CP$ with three kahler
forms $\bomega$ has a symmetry group $G$ generated by
tri-holomorphic vector fields $V^a$, and so $\CL_{V^a} \bomega=
d(i_{V^a}\bomega)+i_{V^a}d\bomega = 0$. As $d\bomega=0$, there
exists three moment map $\bmu^n$ such that
\be i_{V^a} \bomega \,\,=\,\,d\bmu^a\, . \ee
The hyperkahler quotient space $M=\bmu^{-1}/G$ is again
hyperkahler with induced three kahler forms.

Our objective in this subsection is to express the volume of the
quotient $\CM$ in terms of some integration over the ambient space
$\CP$, which we treat as the flat hyperkahler space
$\CP=\mathbb{H}^n \cong \mathbb{C}^{2n}$. In this work, we treat
the quotient group to be a single abelian group, but the
generalization to non-abelian groups is similar. Our work here is
a straightforward application of the method in
Ref.~\cite{Moore:1997dj}.

We can pick any kahler form out of $\bomega$, say $\omega(x)\equiv
\omega^3=\omega_{\mu\nu}(x)dx^\mu\wedge dx^\nu/2$, to define the
volume of a hyperkahler manifold $\CP$,
\be vol(\CP)=\int_{\CP}\,e^{\omega}=\frac{1}{ (2n)!} \int
_{\CP}\,\omega^{2n}\, ,\label{vol} \ee
where ${\rm dim}_{C}\CP=2n$. The normalization at this point is
obscure, but later we will fix it to reproduce the flat volume of
$\mathbb{H}^n\cong \mathbb{C}^{2n}$. Fixing normalization for the
ambient space then unambiguously determine that of the quotient
space.

We introduce mutually anti-commuting Grassmann variables
$\psi^\mu$ which replace the 1-form variable $dx^\mu$, and rewrite
the kahler form as $\omega(x,\psi)
=\omega_{\mu\nu}(x)\psi^\mu\psi^\nu/2$, which is a function of a
bosonic coordinates $x^\mu$ and fermionic variables $\psi^\mu$.
Any differential form $f$ in the space of differential forms,
$\Omega^*(\CP)$, can be regard as a function $f(x,\psi)$. We can
consider $(x,\psi)$ as parameterizing a supermanifold $\CP'$, in
which usual tangent space is fermionic rather than bosonic, say
$\psi^\mu \partial/\partial x^\mu$. With this notation, the
integration of a top differential form $f$ on $\CP$ is written in
a way that mimics supersymmetric functional integration,
\be \int_{\CP}\,f = \int_{\CP'} \,dx^1dx^2\cdots dx^{4n}d\psi^1
d\psi^2 \cdots d\psi^{4n} \,f(x,\psi)= \int_{\CP'} \,[dx][d\psi]
\, f(x,\psi)\, . \ee
It can be checked by calculating super-Jacobian that the measure
$[dx][d\psi]$ is invariant under coordinate reparametrization, and
so is well-defined everywhere on $\CP$. The volume formula in
(\ref{vol}) is then written
\be vol(\CP)=\int_{\CP'} \,[dx][d\psi]\, e^{\omega}\, . \ee
Note that $\int \,[dx][d\psi]$ automatically picks up the top
dimensional form in the expansion of $e^{\omega}$ due to the
properties of the Grassmann integration.

Now let us consider the volume of the hyperkahler quotient space
$\CM$ of the flat space $\CP$ by a $U(1)$ action, which is
generated by the Killing vector $V=V^\mu (x)\partial/\partial
x^\mu$ which preserves the triplet of kahler forms,
\be \CL_V \bomega =0 \,\,\longrightarrow\,\, i_V \bomega =d\bmu \,
. \ee
The hyperkahler quotient space $\CM$ is defined as
$\CM=\bmu^{-1}(0)/U(1)$, which is again hyperkahler with kahler
forms naturally inherited from $\CP$.

To describe the quotient procedure more explicitly, we first note
that $\CL_V \bmu= d\bmu (V)=i_V \bomega (V)=\bomega(V,V)=0$, which
shows that the level surface $\bmu^{-1}(0)$ is invariant under
$U(1)$ flow. Then the Killing vector $V^\mu$ is parallel to
$\bmu^{-1}(0)$. We can therefore introduce a local coordinate
system $(x^i,x^v)$ on (4n-3)-dim space $\bmu^{-1}(0)$ such that
$V=\partial /
\partial x^v$, and $x^i$ ($i=1,\ldots,4n-4$) are constant along
$U(1)$ trajectories. As we further quotient along the direction of
$V$ to get $\CM$, we naturally identify $x^i$ as a coordinate
system on $\CM$. In the ambient space $\CP$, $\bmu^{-1}(0)$ is
codimension 3, so we locally introduce $x^\ell$ ($\ell=1,2,3$)
around $\bmu^{-1}(0)$ as the coordinates along normal directions.
In the following, we only consider points on $\bmu^{-1}(0)$ ( or
$x^\ell=0$) unless stated otherwise. In components, the equation
$i_V \bomega=d\bmu$ becomes
\be \bomega_{vi} dx^i +\bomega_{vn}dx^n = {\partial \bmu \over
\partial x^i}dx^i +{\partial \bmu \over \partial
x^\ell}dx^\ell=\partial_i\bmu \,dx^i+\partial_\ell\bmu \,dx^\ell\,
. \ee
Because $\bmu=0$ on $\bmu^{-1}(0)$, we have $\partial_i \bmu =0$,
and we get
\be \bomega_{vi}=0\, ,\qquad \bomega_{v\ell}=\partial_\ell\bmu\,
.\label{normal} \ee
 From the closed-ness equation $d\bomega=0$, we have $\partial_v
\bomega_{ij}=
\partial_i\bomega_{vj}-\partial_j\bomega_{vi}$, but
$\bomega_{vi}=0$ on every $\bmu^{-1}(0)$ and its tangent
derivatives also vanish, so that $\partial_v \bomega_{ij}=0$. This
means that $\bomega_\CM=\bomega_{ij}dx^i\wedge dx^j/2$ does not
vary along $V$, and so is well defined on $\CM$, and so is
identified as the induced triplet kahler forms.

As we have identified a coordinate system and triplet kahler forms
on the quotient space $\CM$, its symplectic volume will be
\be vol(\CM)=\int_{\CM'}\,[dx^i][d\psi^i]\,e^{{1\over
2}\omega_{ij}\psi^i\psi^j}\, ,\label{quovol} \ee
where $\omega_{ij}=\omega^3_{ij}(x^i)$ on $\bmu^{-1}(0)$ and
$\CM'$ is the corresponding supermanifold for $\CM$. Our aim is to
rewrite (\ref{quovol}) as an integration over the ambient space
$\CP$, which doesn't depend manifestly on a particular coordinate
system we have chosen in the above. Firstly, because $\omega_{ij}$
is independent of $x^v$, the integral can be extended to
$\bmu^{-1}(0)$ as
\be \int_{\CM'}\,[dx^i][d\psi^i]\,e^{{1\over
2}\omega_{ij}\psi^i\psi^j}={1\over vol(U(1))}\int_{\bmu^{-1}(0)}\,
[dx^i][dx^v][d\psi^i]\,e^{{1\over 2}\omega_{ij}\psi^i\psi^j}\, ,
\ee
where $vol(U(1))=\int dx^v$ is the range of the coordinate $x^v$.
To confine the integration on $\CP$ onto $\bmu^{-1}(0)=\{x^n=0\}$,
we would need a $\delta$-function factor $\prod_{a=1}^3
\delta(\mu^a(x))$, and to correctly reduce $\int
\,[dx^i][dx^v][dx^n]$ into $\int \,[dx^i][dx^v]$, we have to add a
Jacobian factor,
\ba \int_{ \bmu^{-1}(0)}\,[dx^i][dx^v]&=&\int_\CP
\,[dx^i][dx^v][dx^\ell]\, \prod_{a=1}^3
\delta(\mu^a(x))\det\left(\partial_\ell
\mu^a\right)\nonumber\\
& & \!\! \!\!\!\!\!\! \!\!\! \!\!\!\! ={1\over (2\pi)^3} \int_\CP
\,[dx^i][dx^v][dx^n][d\psi^\ell][d\phi_a][d\chi_a] \,e^{i\phi_a
\mu^a +\chi_a (\partial_\ell \mu^a) \psi^\ell}\; , \ea
where $\phi_a$ are bosonic and $\chi_a,\psi^\ell$ are fermionic
variables. From (\ref{normal}), we have $\partial_\ell
\mu^3=\omega^3_{v\ell}=\omega_{v\ell}$, and calling
$\chi_3\equiv\psi^v$, we have $\chi_3 (\partial_\ell \mu^3)
\psi^n= \omega_{v\ell}\psi^v\psi^\ell =
\omega_{v\mu}\psi^v\psi^\mu$ since $\omega_{vi}=0$ on
$\bmu^{-1}(0)$. Similarly
$\chi_1\partial_\ell\mu^1\psi^\ell+\chi_2\partial_\ell\mu^2\psi^\ell
=\chi_1 \partial_\mu \mu^1\psi^\mu+ \chi_2\partial_\mu
\mu^2\psi^\mu \equiv \chi_1d\mu^1+\chi_2d\mu^2$ since $
\partial_v\mu^a=0, \partial_i\mu^a=0$ on $\bmu^{-1}(0)$.
Therefore, the volume of $\CM$ is written as
\ba vol(\CM) = {1\over (2\pi)^3 \,vol(U(1))} \int_{\CP''} &&
[dx^i][dx^v][dx^n][d\psi^i][d\psi^v][d\psi^n][d\phi_a][d\chi_{1,2}]
\nonumber\\&\times&e^{{1\over 2}\omega_{ij}\psi^i\psi^j+
\omega_{v\mu}\psi^v\psi^\mu +i\phi_a \mu^a +\chi_1 d\mu^1+\chi_2
d\mu^2}\, . \ea
where $\CP''$ is the space a bit bigger than the supermanifold
with additional coordinates $\phi_a, \chi_{1,2}$.  The terms in
the exponent involving $\omega$ almost comprise the kahler form
${1\over 2}\omega_{\mu\nu}\psi^\mu\psi^\nu$ on X, except
$\omega_{in}\psi^i\psi^n+{1\over 2}\omega_{mn}\psi^m\psi^n$.
However, the first term can be removed by shifting $\psi^i$, and
also the second term by shifting $\chi_a$. Therefore, we can
replace the exponent by ${1\over
2}\omega_{\mu\nu}\psi^\mu\psi^\nu+i\phi_a \mu^a +\chi_1
d\mu^1+\chi_2 d\mu^2$ without changing the result. Finally, the
measure involves $[dx][d\psi]$ on $X$, which is manifestly
independent of a particular coordinate choice. Thus, we have
\be vol(\CM)= {1\over (2\pi)^3\, vol(U(1))}
\int_{\CP''}\,[dx][d\psi][d\phi_a][d\chi_{1,2}] \;\; e^S \, ,\ee
where the `action' is
\ba && S(x,\psi, \phi_a,\chi_1,\chi_2) =\omega +i\phi_a \mu^a
+\chi_1 d\mu^1+\chi_2 d\mu^2 \nonumber \\
&& \;\;\;\;\;\;\;\;\; =
\frac{1}{2}\omega_{\mu\nu}(x)\psi^\mu\psi^\nu +i\phi_a\mu^a(x) +
\chi_1\partial_\mu\mu^1(x)\psi^\mu+\chi_2\partial_\mu\mu^2(x)\psi^\mu
\, . \ea

The above integration looks like a path integral of a
(0+0)-dimensional supersymmetric system. Indeed, the action $S$
has the following fermionic symmetry,
\ba && Q\, x^\mu = \psi^\mu\, , \, \;\;
Q\,  \psi^\mu = -i\phi_3 V^\mu(x)\, ,  \nonumber\\
&& Q\,\phi_a = 0\, ,\;\; Q\, \chi_1 = -i\phi_1 \, ,\;\; Q\, \chi_2
= -i\phi_2\, , \label{defineQ}\ea
which can be easily verified using $d\omega=0$ and $i_V
\bomega=d\bmu$. Acting $Q$ twice, we have $Q^2\, x^\mu=-i\phi_3
V^\mu$ and $Q^2 \,\psi^\mu=-i\phi_3(\partial_\nu V^\mu)\psi^\nu$,
while $Q^2\,(\phi,\chi) =0$. On the space of functions on
$\Omega^*(\CP)$, that is, on the space of differential forms on
$X$, this is nothing but $Q^2 = -i \phi_3 \CL_V$. Therefore, if we
restrict to the space of $U(1)$-invariant differential forms, $Q$
is nilpotent. In fact, the right observables well-defined on $\CM$
are indeed $U(1)$-invariant differential forms on $\CP$, and the
correlation functions of them depend only on the $Q$-cohomology.
These correlation functions are nothing but the integrals on $\CM$
performed in the ambient space $\CP$.

Suppose that there is a $U(1)_R$-action on $\CP$ generated by
$R^\mu(x)$ such that $\CL_R \omega=\CL_R\omega^3=0$, $\CL_R
(\omega^1-i\omega^2)=2i(\omega^1-i\omega^2)$, which imply that
$\CL_R \mu^3=R^\alpha\partial_\alpha \mu^3=0$, and $\CL_R
(\mu^1-i\mu^2)=R^\alpha\partial_\alpha (\mu^1-i\mu^2)
=2i(\mu^1-i\mu^2)$. The former implies that there is a function
$H(x)$ with $i_R \omega=dH$. We also assume that $V$ commutes with
$R$, that is, $[V,R]=0$. Then we naturally assign the $R$-action
to $\phi_a$ and $\chi_{1,2}$ such that the integrand $S$ respects
this symmetry;
\be R \cdot(\phi_1-i\phi_2)= 2i(\phi_1-i\phi_2)\, ,\;\;  R
\cdot(\chi_1-i\chi_2) = 2i(\chi_1-i\chi_2)\, , \ee
and $R \cdot\phi_3 =0$. This allows us to deform the supersymmetry
$Q$ to $Q_\epsilon$ in the following way,
\ba && Q_\epsilon\, x^\mu = \psi^\mu\, , \; \; Q_\epsilon\,
\psi^\mu =-i\phi_3 V^\mu(x)+\epsilon R^\mu(x)\, ,\;\;
Q_\epsilon\,\phi_3 = 0\, , \nonumber\\
&& Q_\epsilon\,\phi_1 = 2i\epsilon\chi_2\, ,\;\;
Q_\epsilon\,\phi_2 = -2i\epsilon\chi_1\, ,\;\; Q_\epsilon\,
\chi_{1} = -i\phi_{1}\,  ,\;\; Q_\epsilon\, \chi_{2}= -i\phi_{2}\,
, \label{defineQe}\ea
with $Q_\epsilon^2=-i\phi_3\CL_V+\epsilon R$, where $R$ acts as
$\CL_R$ on differential forms, and $\epsilon$ is a constant. Thus,
$Q_\epsilon^2=0$ still holds on the space of both $U(1)$ and
$U(1)_R$ invariant functions. It is straightforward to check that
$S$ is both $U(1)$ and $U(1)_R$ invariant, and
\be Q_\epsilon \,S=\epsilon \,\omega_{\mu\nu}
R^\mu(x)\psi^\nu=\epsilon \,i_R\omega = \epsilon\, dH=Q_\epsilon
\left(\epsilon H(x)\right) \, . \ee
Hence, $S_\epsilon\equiv S-\epsilon H$ is $Q_\epsilon$-invariant.
For non-compact hyperkahler manifolds, the term $-\epsilon H$
often provides a regularization for volume \cite{Moore:1997dj}. In
addition, the regularized volume integration would be left
unchanged if we add a bosonic $Q_\epsilon$-exact term $Q_\epsilon
\cal{O}$ to the deformed action $S_\epsilon$, where fermionic
$\cal{O}$ should be $U(1)$ and $U(1)_R$-invariant, so that
$Q_\epsilon^2 {\cal{O}}=0$. We then consider the regularized
volume of $\CM$,
\be vol_\epsilon(\CM)={1\over (2\pi)^3\, vol(U(1))} \int_{\CP''}
\,[dx][d\psi][d\phi_a][d\chi_{1,2}] \,\,e^{\,S-\epsilon H + \,
Q_\epsilon \cal{O}'}\, . \label{regvolume}\ee
One useful $Q_\epsilon \cal{O}'$ is
\be Q_\epsilon {\cal{O}'} = Q_\epsilon \biggl(-it
(\chi_1\phi_1+\chi_2\phi_2)\biggr)
=-t(\phi_1\phi_1+\phi_2\phi_2)-4\epsilon t (\chi_1\chi_2)\,
,\label{add} \ee
which will dominate the $\phi_{1,2}$ and $\chi_{1,2}$ terms in the
action when we take $t\to\infty$ limit, which allows a simple
integration over $\phi_{1,2}, \chi_{1,2}$. The remaining
integration over $x^\mu$, $\psi^\mu$ and $\phi_3$ will then be
simple Gaussian in our case and can be performed easily.

\noindent{\it Calculations for} $ X_7(\bt)$

We now apply the previous formalism to an explicit problem of our
space $X_7(\bt)=S(t_1,t_2,t_3)$. We started with a hyperk\"ahler
quotient of the flat 12-dim hyperkahler space
$\CP=\mathbb{H}^3=(q_1,q_2,q_3)$. We obtained the 8-dim
hyperkahler space $\CM$ by the hyperkahler quotient of $\CP$ by
 a $U(1)$ action $ q_a \rightarrow q_ae^{i\sigma_3 t_a \xi }$,
$a=1,2,3$. Here we use the representation of the quaternions $q_a$
by the complex coordinates $u_a, v_a$ as in Eq.(\ref{uvcoord}),
which in turn be represented by the real coordiantes
\be q_a = \left(
\begin{array}{rr} u_a & v_a \\ -\bar{v}_a & \bar{u}_a
\end{array}\right) = \left(
\begin{array}{rr} x_a+iy_a &\;\; \tilde{x}_a+i\tilde{y}_a \\
-\tilde{x}_a+i\tilde{y}_a &\;\; x_a-iy_a
\end{array}\right)\, . \ee
The triplet hyperkahler forms (\ref{kahlerforms}) become
\ba && \omega^3=-(dx_a\wedge dy_a +d\tilde{x}_a\wedge d\tilde{y}_a
)\, ,
\nonumber \\
&& \omega^1-i\omega^2= i(dx_a\wedge d\tilde{x}_a -dy_a\wedge
d\tilde{y}_a) -(dx_a\wedge d\tilde{y}_a+dy_a\wedge d\tilde{x}_a)
\, . \label{omega}\ea
With these canonical coordinates, the volume $\int
[dx][d\psi]\,e^{ \omega}$ of the ambient space $\CP=\mathbb{H}^3$
is simply the flat volume $\int [dx_a][dy_a][d\tilde x_a][d\tilde
y_a]$. This fixes the normalization of the ambient space metric to
be $ds_{\mathbb{H}^3}^2=dx_a dx_a+dy_a dy_a+d\tilde x_a d\tilde
x_a+d\tilde y_a d\tilde y_a$. In components, the above $U(1)$
action is
\be u_a\to e^{it_a \xi} u_a\, ,\quad v_a\to e^{-i t_a\xi }v_a\, ,
\ee
so that the generating vector field $V$ is
\be V={\partial\over\partial\xi}=t_a\left(x_a
{\partial\over\partial y_a}-y_a {\partial\over\partial x_a}\right)
-t_a\left(\tilde x_a {\partial\over\partial \tilde y_a}-\tilde y_a
{\partial\over\partial \tilde x_a}\right) \, . \ee
 From the definition of $i_V\bomega=d\bmu$, we have
\be \mu^3={1\over 2} t_a\left(|u_a|^2 -|v_a|^2\right)\, ,\quad
\mu^1-i\mu^2= -t_a u_a v_a\, . \ee
In addition, there is a diagonal $U(1)_R$ action of $SU(2)_R$ with
the fore-mentioned properties,
\be u_a\to e^{i\epsilon} u_a\, ,\quad v_a\to e^{i\epsilon}v_a\, ,
\ee
which gives us $R$ as
\be R=\left(x_a {\partial\over\partial y_a}-y_a
{\partial\over\partial x_a} +\tilde x_a {\partial\over\partial
\tilde y_a}-\tilde y_a {\partial\over\partial \tilde x_a}\right)
\, , \ee
and from $i_R\omega=dH$, we have $H={1\over
2}(|u_a|^2+|v_a|^2)={1\over 2} r^2$ with $r$ being the standard
radial distance in the above flat metric.

The volume of $U(1)$ is the coordinate length of $\xi$, that is,
the least number $\xi$ such that $t_a \xi\in 2\pi\mathbb{Z}$ for
all $a=1,2,3$. It is easily seen to be
\be vol(U(1))=(2\pi){ l.c.m}\left({1\over t_a}\right)={2\pi\over
t_1 t_2 t_3} { l.c.m}(t_1t_2,t_2t_3,t_3t_1)\, , \ee
where $l.c.m$ stands for least common multiple.

Let us integrate over $\phi_{1,2}$ and $\chi_{1,2}$ in the
regularized volume integration (\ref{regvolume}) in the large $t$
limit, $\CM_8$. The integration is dominated by $Q_\epsilon {\cal
O}$ to give
\be \int \,d\phi_1 d\phi_2 d\chi_1 d\chi_2\,
e^{-t(\phi_1\phi_1+\phi_2\phi_2)-4\epsilon t (\chi_1\chi_2)}
={\pi\over t}\cdot 4\epsilon t = 4\pi\epsilon\, . \ee
The remaining integration is
\ba vol_\epsilon({\cal M}_8(\bt))= {4\pi\epsilon\over (2\pi)^3\,
vol(U(1))} \int_{\mathbb{H}^3}\,[dx][d\psi]d\phi_3 \,\,e^{{1\over
2}\omega_{\mu\nu}\psi^\mu\psi^\nu+i\phi_3\mu^3(x) -\epsilon
H(x)}\, . \ea
Note that $\omega$ in (\ref{omega}) is simply constant in the flat
coordinate $(x_a,y_a,\tilde x_a,\tilde y_a)$ of $\CP$, and the
$[d\psi]$ integration readily calculated to be
\be \int\,[d\psi] \,e^{{1\over
2}\omega_{\mu\nu}\psi^\mu\psi^\nu}=1\, . \ee
Also, $\mu^3$ and $H$ are both Gaussian functions on $x^\mu$ and a
simple calculation gives
\ba \int_{\mathbb{H}^3}\,[dx]\, e^{i\phi_3 \mu^3 -\epsilon H}&=&
\int\,[dx]\, e^{{i\over 2}\phi_3t_a\left((x_a)^2+(y_a)^2-(\tilde
x_a)^2 -(\tilde y_a)^2\right) -{\epsilon\over 2}\left(
(x_a)^2+(y_a)^2+(\tilde x_a)^2 +(\tilde
y_a)^2\right)}\nonumber\\
&=&(2\pi)^6 \prod_{a=1}^3\,{1\over (\epsilon-it_a
\phi_3)(\epsilon+it_a \phi_3)}\, , \ea
so that
\be vol_\epsilon({\cal M}_8(\bt))= {4\pi\epsilon(2\pi)^3\over
vol(U(1))}\int\,d\phi_3\,\prod_{a=1}^3\,{1\over (\epsilon-it_a
\phi_3)(\epsilon+it_a \phi_3)} \, .\ee
The $\phi_3$ integration has poles at $\phi_3=\pm {i\epsilon\over
t_a}$, and by closing the contour to the upper half plane, we pick
up poles at $\phi_3= {i\epsilon\over t_a}$, $a=1,2,3$. The result
is
\be \int\,d\phi_3\,\prod_{a=1}^3\,{1\over (\epsilon-it_a
\phi_3)(\epsilon+it_a \phi_3)} = {\pi\over
\epsilon^5}\,{t_1t_2+t_2t_3+t_3t_1 \over
(t_1+t_2)(t_2+t_3)(t_3+t_1)}\, . \ee
In summary, we have the regularized volume of the 8-dim quotient
space
\be vol_\epsilon({\cal M}_8(\bt))={16\pi^4 \over \epsilon^4}
{t_1t_2t_3(t_1t_2+t_2t_3+t_3t_1) \over {
l.c.m}(t_1t_2,t_2t_3,t_3t_1) (t_1+t_2)(t_2+t_3)(t_3+t_1)}\, . \ee
While we started with three distinct natural numbers $t_a$, the
above formula is  well defined for any postive real number $t_a$
and so can be regarded valid even when some of $t_a$ coincide.
(Note the above procedure can be easily generalized to the
regularized volume for the hyperkahler quotient space $\CM_{4n}$
obtained from the $\mathbb{H}^{n+1} $ by a single $U(1)$.)

As our objective is to calculate the volumes of 7-dim tri-Sasakian
section of 8-dimensional hyperkahler cones, this Hamiltonian
regularization happens to be exactly what we would need to extract
the volumes of tri-Sasakian section. This is because, in our cases
at hand, $H$ will turn out to be $H={1\over 2}r^2$, when the
metric is written as $ds_8^2 = dr^2 + r^2 ds_{X_7}^2$, and the
regularized volume is \be vol_\epsilon({\cal M}_8 ) =
vol(X_7)\,\int_0^\infty r^7 e^{-{1\over 2}\epsilon r^2}={48\over
\epsilon^4}vol(X_7)\quad.\label{x7} \ee Finding the regularized
volume would give us the volume of the tri-Sasakian section with
normalization $R_{ij}= 6 g_{ij}$.

By using (\ref{x7}), and we obtain the formula
\be \frac{vol(X_7(\bt))}{vol(S^7) } =
{t_1t_2t_3(t_1t_2+t_2t_3+t_3t_1) \over {
l.c.m}(t_1t_2,t_2t_3,t_3t_1) (t_1+t_2)(t_2+t_3)(t_3+t_1)}\, . \ee
The volume of the unit 7 sphere is $vol(S^7)=\pi^4/3$. The ratio
of two volumes is a rational number. As a check, the well-known
space $N(1,1)$ corresponds to $t_1=t_2=t_3=1$ for which we
reproduce the known answer $vol(N(1,1))={\pi^4\over 8}$. We can
show that the right hand side is less than $3/8$. The reason is
that for three natural numbers $t_1,t_2,t_3$ the following
inequalties hold,
\ba && t_1t_2+t_2t_3+t_3t_1 \le 3\;{l.c.m}(t_1t_2,t_2t_3,t_3t_1),
\nonumber \\
&& t_1t_2t_3 \le \frac{1}{8}(t_1+t_2)(t_2+t_3)(t_3+t_1)\, , \ea
where the last inequality comes from $2\sqrt{t_1t_2}\le t_1+t_2$
and so on. The equality holds only when $t_1=t_2=t_3$.

One could study the tri-Sasakian space $\CM_{4n-1}$ obtained from
the hyperkahler quotient of $\mathbb{H}^{n+1}$ by a similar
$U(1)_\bt$ group. The volume can be calculated by the above method
and is equal to
\be \frac{vol(M_{4n-1})}{vol(S_{4n-1})} =
\frac{1}{l.c.m.(1/t_1,...,1/t_{n+1})} \sum_a
\frac{t_a^{2n-1}}{\prod_{b\neq a}(t_a^2-t_b^2)}\, , \ee
where the volume of the unit sphere $vol(S_{2n-1})$ is $(2n-1)! \;
\pi^{n}/2 $.

\subsection{The volumes of supersymmetric 5-cycle $\Sigma_5$ }

The cone of the tri-Sasakian spaces $X_7=S(t_1,t_2,t_3)$ are
8-dimensional hyperk\"ahler spaces, which are constructed through
hyperkahler quotient. A supersymmetric 5-cycle, $\Sigma_5$ in
$S(t_1,t_2,t_3)$ is characterized by its cone $\Gamma$, which is a
6-dimensional subspace of the hyperk\"ahler cone $\CM(\bt)$,
defined by a single homogeneous holomorphic constraint. In fact,
the two 5-cycles in $N(1,1)=S(1,1,1)$ that were identified in
Ref.\cite{Gauntlett:2005jb} correspond to $u_3=0$ and $v_3=0$
respectively. Of course, there are others such as $u_1=0$ et
cetera, which are related to each other by continuous $SU(3)$
isometry. They necessarily belong to the same homology. For
generic $S(t_1,t_2,t_3)$, the constraint $u_a=0$ or $v_a=0$ for
some $a$ again defines a supersymmetric 5-cycle, but the flavor
isometry is now reduced to $U(1)^2$ and the cycles with different
$a$'s are separated by potential walls. The remaining $SU(2)_R$
symmetry still relate $u_a=0$ and $v_a=0$ for the same $a$.

In this section, we calculate the volumes of supersymmetric
5-cycles using the formalism of the previous section. Without loss
of generality, we specify to a supersymmetric 5-cycle $\Sigma_5$
obtained from the constraint, say with $u_3=0$, whose 6-dim cone
is $\Gamma$. Before considering $\Gamma$ in our hyperkahler
quotient space, let us consider the 10-cycle $\tilde\Gamma$
defined in the ambient space $\mathbb{H}^3$ by the same constraint
$u_3=0$. Its volume, though infinite, is expressed formally as
\be {1\over 5!}\int_{u_3=0}\,\omega^5={1\over
5!}\int_{\mathbb{H}^3}\, \Phi_{\tilde\Gamma}\wedge \omega^5\, ,
\ee
where $\omega$ is the Kahler form, and $\Phi_{\tilde\Gamma}$ is
the 2-form Thom class dual to $\tilde\Gamma$. In the real
coordinate system introduced in the last section, $u_3=x_3+iy_3$
and $\Phi_{\tilde\Gamma}=\delta(x_3)\delta(y_3) dx_3\wedge dy_3$.
It is easily verified that $d\Phi_{\tilde\Gamma}=0$. In the
formalism of the previous section where differential forms are
functions on $T_{[1]}X$, we can rewrite the above as an
expectation value,
\be vol(\tilde\Gamma)=\langle
\Phi_{\tilde\Gamma}\rangle=\int_{\mathbb{H}^{3'}}
\,[dx][d\psi]\,\Phi_{\tilde\Gamma}\,e^{ \omega}\, , \ee
where
$\Phi_{\tilde\Gamma}=\delta(x_3)\delta(y_3)\psi^{x_3}\psi^{y_3}$
and $\omega=\omega_{\mu\nu}\psi^\mu\psi^\nu/2$. The `action'
$S=\omega$ is invariant under a fermionic symmetry $Q
x^\mu=\psi^\mu$, $Q\psi^\mu=0$ due to $d\omega=0$ ($Q$ is in fact
the de Rham d-operator on differential forms). Because
$Q\Phi_{\tilde\Gamma}=0$ ($d\Phi_{\tilde\Gamma}=0$),
$\Phi_{\tilde\Gamma}$ is a good observable of the above path
integral.

For our quotient space ${\cal M}_8(\bt)$, we have represented the
regularized volume as a path integral
\be vol_\epsilon({\cal M}_8(\bt))=\langle \,1\,\rangle_\epsilon
={1\over (2\pi)^3\, vol(U(1))}
\int\,[dx][d\psi][d\phi_a][d\chi_{1,2}] \,\,e^{\,S-\epsilon H+
Q_\epsilon{\cal O}' }\, . \ee
It also has a fermionic symmetry $Q_\epsilon$ as discussed before.
Actually, there are two separate components in $S_\epsilon$ that
are $Q_\epsilon$-invariant; one is ${1\over 2}\omega+i\phi_3 \mu^3
-\epsilon H$, the other being $i\phi_1 \mu^1 +i\phi_2 \mu^2+\chi_1
d\mu^1+\chi_2 d\mu^2$. In the spirit of equivariant cohomology, a
good observable $\cal{O}$ satisfying $Q_\epsilon \cal{O}$
represents a well-defined geometric data on the quotient space. In
this respect, the first piece may be considered as the Kahler form
of the quotient space, and our path integral naturally calculates
the regularized volume of the quotient space.

Taking this analogy further, we expect there should exist the
right observable $\Phi_{\Gamma}$ whose expectation value
calculates the regularized volume of $\Gamma$ defined by $u_3=0$
in the quotient space. In the ambient space $\mathbb{H}^3$, it is
$\Phi_{\tilde\Gamma}$ and we naturally expect that for quotient
space, it would be some modification of $\Phi_{\tilde\Gamma}$ such
that it is $Q_\epsilon$-invariant. In our case at hand, it is
readily shown that $\Phi_{\tilde\Gamma}$ itself is
$Q_\epsilon$-invariant because of $\delta$-function factors.
Therefore, we take $\Phi_{\Gamma}=\Phi_{\tilde\Gamma}$, and the
regularized volume of $\Gamma$ will be
\be vol_\epsilon(\Gamma)=\langle \Phi_{\Gamma}\rangle_\epsilon
={1\over (2\pi)^3\, vol(U(1))}
\int\,[dx][d\psi][d\phi_a][d\chi_{1,2}]
\,\,\Phi_{\Gamma}\,\,e^{\,S-\epsilon H+Q_\epsilon{\cal O}'}\,
.\label{regvol} \ee
{}From $vol_\epsilon(\Gamma)$, it is straightforward to extract
the volume of 5-dimensional cycle $\Sigma_5$ that we are heading
to. Writing the metric on $\Gamma$ as $ds^2_\Gamma=dr^2 +r^2
ds^2_{\Sigma_5}$,
\be vol_\epsilon(\Gamma)=vol(\Sigma_5)\,\int_0^\infty
dr\,r^5\,e^{-{1\over 2}\epsilon r^2} =vol(\Sigma_5)\cdot{8\over
\epsilon^3}\, . \ee

The calculation of (\ref{regvol}) is almost same as the one in the
previous section. Introducing large $Q_\epsilon$-exact mass term
(\ref{add}) for $\phi^{1,2}$ and $\chi^{1,2}$, and integrating
them out results in
\be vol_\epsilon(\Gamma)={4\pi\epsilon\over (2\pi)^3 vol(U(1))}
\int_{\mathbb{H}^3}\,[dx][d\psi]d\phi_3\,\,\delta(x_3)\delta(y_3)\psi^{x_3}\psi^{y_3}
\,\,e^{\omega+i\phi^3\mu^3(x) -\epsilon H(x)}\, . \ee
By using simple integrations
\ba && \int\,[d\psi] \,\psi^{x_3}\psi^{y_3}\,e^{{1\over
2}\omega_{\mu\nu}\psi^\mu\psi^\nu}= 1 \, , \nonumber \\
&& \!\!\!\!\!\!\!\!\!\!\!
\int_{\mathbb{H}^3}\,[dx]\,\,\delta(x_3)\delta(y_3)\, e^{i\phi^3
\mu^3 -\epsilon H} =(2\pi)^5 \,{1\over (\epsilon-it_1
\phi^3)(\epsilon-it_2 \phi^3)\prod_{a=1}^3(\epsilon+it_a
\phi^3)}\, , \ea
we are left with
\be vol_\epsilon(\Gamma)={4\pi\epsilon (2\pi)^2 \over
vol(U(1))}\int d\phi^3\,\, {1\over (\epsilon-it_1
\phi^3)(\epsilon-it_2 \phi^3)\prod_{a=1}^3(\epsilon+it_a
\phi^3)}\, . \ee
The integrand has poles at $ -{i\epsilon\over
t_1},-{i\epsilon\over t_2}$ and $+{i\epsilon\over t_a}, a=1,2,3$.
Because it is convergent, we can close the contour in any way, and
the result is
\be \int d\phi^3\,\, {1\over (\epsilon-it_1 \phi^3)(\epsilon-it_2
\phi^3)\prod_{a=1}^3(\epsilon+it_a \phi^3)} ={\pi\over
\epsilon^4}\cdot{t_1t_2+t_2t_3+t_3t_1\over
(t_1+t_2)(t_2+t_3)(t_3+t_1)}\, , \ee
so that
\be vol_\epsilon(\Gamma)={16\pi^4\over vol(U(1))}
{t_1t_2+t_2t_3+t_3t_1\over
(t_1+t_2)(t_2+t_3)(t_3+t_1)}\cdot{1\over \epsilon^3}
=vol(\Sigma_5)\cdot{8\over \epsilon^3}\, , \ee
and we finally obtain
\be \frac{vol(\Sigma_5)}{vol(S^5)}=
{t_1t_2t_3(t_1t_2+t_2t_3+t_3t_1) \over {
l.c.m}(t_1t_2,t_2t_3,t_3t_1) (t_1+t_2)(t_2+t_3)(t_3+t_1)}\, , \ee
where the volume of unit five sphere is $vol(S^5)=\pi^3$. The
above expression is identical to $vol(X_7)/vol(S^7)$. For
$t_1=t_2=t_3=1$, it is ${3\pi^3\over 8}$ which agrees with the
known value of supersymmetric 5-cycles of $N(1,1)$ in
Ref.\cite{Gauntlett:2005jb}.

Interestingly, the 5-cycles $u_a=0$ or $v_a=0$ have the same
volume independent of $a$. The more striking fact is that the
volume ratio between the supersymmetric 5-cycle and the total
tri-Sasakian 7-manifold is independent of $t_a$ and takes a
universal value
\be {vol(\Sigma_5)\over
vol(X_7(\bt))}=\frac{vol(S^5)}{vol(S^7)}={3\over \pi}\, .
\label{ratio}\ee
This will turn out to be consistent with the SCFT expectation.

\section{Dual Superconformal Field Theory}

The bosonic Lagrangian for 11-dim supergravity is
\be 2\kappa^2 {\cal L} = \sqrt{-G}{\cal R} - \frac{1}{2} F_4\wedge
*F_4 -\frac{1}{6}C_3\wedge F_4\wedge F_4 \, , \ee
in the convention of Ref.\cite{Gauntlett:2005jb}. The 11-dim
Planck length $\l_{11}$ would be given by $2\kappa^2 =
(2\pi)^8\l_{11}^9$. In this convention, the $M2$ charge and $M5$
charges are given by the flux of $F_4$ as
\be N_2 = \frac{1}{(2\pi \l_{11})^6}\int_{C_7} *F_4\, , \;\; N_5 =
\frac{1}{(2\pi \l_{11})^3}\int_{C_4}F_4 \, , \label{m2m5} \ee
for some 7-cycle $C_7$ and 4-cycle $C_4$ surrounding branes. Our
supersymmetric solution of the supergravity takes the form
\be ds_{11}^2 = R_{X_7}^2
\biggl(\frac{1}{4}ds^2_{AdS_4}+ds_{X_7}^2\biggr)\, , \;\; F_4 =
\frac{3}{8}R_{X_7}^3 \; vol_{AdS_4} \, ,\ee
with $vol_{AdS_4}$ being the volume form of the $AdS_4$ space with
metric $ds_{AdS_4}^2$. Here we assume that the normalization of
$ds_{X_7}^2$ is such that $R_{\mu\nu}(X_7)=6g_{\mu\nu}(X_7) $. The
radius of $R_{X_7}$ is given by the quantization condition of M2
brane charge $N_2$,
\be 6R_{X_7}^6 {vol}(X_7) = (2\pi \l_{11})^6 N_2 \, ,\ee
where ${vol}(X_7)$ is the volume of $X_7$, which we calculate this
volume in the section ahead.

The dual field theory on $M2$ branes on the singular point is a
$\CN=3$ supersymmetric conformal field theory. One naively thinks
that it is the infrared limit of a gauge theory with the product
gauge group
\be SU(N)_1\times SU(N)_2\, , \ee
where $N$ is the M2 charge $N_2$ of Eq.(\ref{m2m5}). In terms of
$\CN=2$ language ($\CN=1$ in 4-dim), there are 6 chiral fields
$u_a, v_a$, $a=1,2,3$. The best way to consider the $SU(2)_R$
symmetry is to group these fields into $U_a^\beta=(u_a,
-\bar{v}_a)$ and $V_{a\beta}=(v_a, \bar{u}_a)$, where $a=1,2,3$
and $\beta=1,2$. The chiral field $U_a^\beta$ belongs to the
symmetrized product representation $Sym^{t_a}(N)$ of the
fundamental representation of the first $SU(N)_1$ and the
symmetrized product representation $Sym^{t_a}(\bar{N})$ of the
second $SU(N)_2$. The chiral field $V_{a\beta}$ transforms
opposite to the chiral field $U_a^\beta$. Under the additional
$U(1)\times U(1)$ global symmetries, the charges are shown in the
following table.
\vspace{1em}

\begin{tabular}{|c|c|c|c|c|c|}
\hline
& $SU(2)_R$ & $SU(N)_1$ & $SU(N)_2$ & $U(1)$ & $U(1)$ \\
\hline $U_a^\beta$ & {\bf 2} & $Sym^{t_a}(N)$ &
$Sym^{t_a}(\bar{N})$ & $(t_1, -t_2, 0)$ & ($0, t_2,-t_3)$ \\
\hline $V_{a\beta}$ & {\bf 2} & $Sym^{t_a}(\bar{N})$ &
$Sym^{t_a}(N)$ & $(-t_1, t_2, 0)$ & $(0,-t_2,t_3)$ \\ \hline
\end{tabular}
\vspace{1em}

As the R-symmetry is nonabelain $SU(2)$ group and these chiral
fields belongs to the fundamental representation of this
R-symmetry, these chiral fields would have a chiral dimension
$1/2$. The chiral operators of dimension $k$ would be, for
example, traces of products of $k$ U's and $k$ V's with totally
symmetric in $SU(2)_R$ indices. There are still many kinds of such
operators with different combinations of flavor indices $a=1,2,3$.
In the case of $N(1,1)=S(1,1,1)$, KK analysis of the geometry
dictates only operators with totally symmetric and traceless in
flavor indices \cite{Fre':1999xp}, and we expect a similar kind of
reduction in the spectrum for generic $S(t_1,t_2,t_3)$. In SCFT,
we have to assume that only these operators survive in the IR
fixed point. As was pointed out in
Ref.\cite{Fabbri:1999hw,Billo:2000zr}, this contrasts to the case
of $AdS_5\times T^{1,1}$ where a superpotential in the dual field
theory selects right chiral primary operators that match with
gravity analysis \cite{Klebanov:1998hh}.

We can try superpotential approach as much as in
\cite{Billo:2000zr}, though it would not be sufficient to
determine chiral primary operators. We introduce the complex
scalar fields $\Phi_1, \Phi_2$ in the adjoint representation of
$SU(N)_1\times SU(N)_2$, respectively. They belong to the vector
multiplet and in component $\Phi_1^i,\Phi_2^i$ with
$i=1,2,...,N^2-1$. We propose the superpotential to be
\be W = \sum_{a=1}^3\biggl( g_1 \Phi_1^i \Tr_a T_a^i
U_{a\beta}V^{a\beta} + g_2 \Phi_2^i \Tr_a T_a^iV_{a\beta}U_a^\beta
\biggr)+ k_1 \Phi_1^i\Phi_1^i + k_2 \Phi_2^i \Phi_2^i \, ,\ee
where $g_1=g_2=g$ are the gauge coupling constants, $T_a^i$ are
matrix representation in $Sym^{t_a}(N)$ and $\Tr_a$ is the trace
operation in this representation. The Chern-Simons coefficients
$k_1$ and $k_2$ may satisfy $k_1=-k_2$ as in the case of
$S(1,1,1)=N(1,1)$ \cite{Billo:2000zr}. Integrating out $\Phi$'s
would produce a superpotential for $U$'s and $V$'s in IR.

The theory has a single $U(1)$ baryon symmetry since the second
and fifth betti numbers are $b_2(S(t_1,t_2,t_3)=
b_5(S(t_1,t_2,t_3))=1$. There are baryonic operators such as ${\rm
det} U_a$ or ${\rm det} V_a$ where det represents the product of
$N$ $U_a$'s or $N$ $V_a$'s totally anti-symmetrized in gauge
indices for both $SU(N)_1$ and $SU(N)_2$. Then $SU(2)_R$ indices
are totally symmetric. The conformal dimension would then be
$\Delta={N \over 2}$ independent of $a$ or $\bt$. From the
geometry, these operators correspond to M5 branes wrapping
supersymmetric 5-cycles with the dimension (\ref{baryonn} and the
result (\ref{ratio}) gives $\Delta={N \over 2}$ which agrees with
the gauge theory expectation. We take this as a non-trivial
evidence for our proposal of superconformal field theory.

One should caution that the detail characteristics of the
corresponding $\CN=3$ SCFT is very obscure. One naively can
imagine that this theory is the low energy conformal limit of the
Yang-Mills-Chern-Simons theory of the gauge group $SU(N)_1\times
SU(N)_2$ and the matter chiral fields $u_a, v_a$. In the low
energy theory massive vector multiplet decouples, and so only the
Chern-Simons kinetic term survives, with matter fields interact
each other by the gauge coupling and self-coupling. The
corresponding t'Hooft coupling is $N/\kappa$ with $\kappa$ being
the Chern-Simons theory coefficient. From our geometric point of
view of $AdS_4\times X_7$, the Chern-Simons coefficient is not
obvious at all. It would be interesting to find out further about
this discrepancy.

\acknowledgments This work is supported in part by KOSEF Grant
R010-2003-000-10391-0 (K.L.,H.U.Y.), KOSEF SRC Program through
CQUeST at Sogang Univ. (K.L.), KRF Grant No. KRF-2005-070-C00030
(K.L.). H.U.Y thanks Sang-Heon Yi and Tetsuji Kimura for helpful
discussion.

\appendix

\section{A Generalization of Caloron Moduli Space}

We start this appendix by considering a generalization of the
moduli space of distinct multi BPS magnetic monopole
solutions\cite{Lee:1996kz}. Instead of the considering the
interaction between two BPS dyonic monopoles whose interactions
are fixed by the Dynkin diagram for a give Lie algebra, we
introduce a somewhat more general interaction between them. Thus
the generalized Lagrangian between multi-monopoles would be
\be L = \frac{1}{2}M_{ij}\left(\dot{\bx}_i\cdot \dot{\bx}_j
-q_iq_j\right)+ q_i\bW_{ij}\cdot\dot{\bx}_j + q_i\dot{\xi}_i\,
,\ee
where
\be M_{ii}=m_i+\sum_{k\neq i} \frac{\lambda_{ik}}{ |\bx_i-\bx_k|},
\;\; M_{ij}=-\frac{ \lambda_{ij}}{ |\bx_i-\bx_j|} \; {\rm if} \;
i\neq j \, ,\ee
with non-negative mass parameters $m_i\ge 0$ and
\be \bW_{ii} = \sum_{k\neq i} \lambda_{ik} \bw(\bx_i-\bx_k),
\;\;\bW_{ij}=-\lambda_{ij}\bw(\bx_i-\bx_j) \; {\rm if} \; i\neq
j\, , \ee
with $\bw$ being the value at $\bx_i$ of the Dirac potential due
to the $j$-th monopole so that
\be \nabla \times \bw(\bx)= \frac{\bx}{|\bx|^3} \, .\ee

The range of each phase is given by
\be 0\le \xi_i< 4\pi t_i \, ,\ee
which implies that the quantization of charge is satisfied with
\be q_i = \frac{n_i}{2t_i} \, \ee
with integer $n_i$. After integration over $q_i$, we obtain the
Lagrangian
\be L = \frac{1}{2}M_{ij} \bv_i\cdot \bv_j +\frac{1}{2}
M^{-1}_{ij}(\dot{\xi}_i +\bW_{ik}\cdot \bv_k ) (\dot{\xi}_j
+\bW_{jl}\cdot \bv_l ) \, .\ee

The geometry is not necessarily regular when two points particles
come together. Now we want the metric to be regular whenever any
of two particles interacting each other come together. This
requires that more detail analysis of two bodies. With the center
of mass coordinate for any two body, say $i=1,2$,
\be \bR = \frac{m_1\bx_1+m_2\bx_2}{m_1+m_2},\;\;\;
\br=\bx_1-\bx_2 \, . \ee
The total charge and relative charge are defined as a linear
combination
\be q_t = \frac{m_1q_1+m_2q_2}{m_1+m_2},\; q_r =
\lambda_{12}(q_1-q_2)\, , \ee
which leads to the c.m. and relative angles
\be \chi=\xi_1+\xi_2, \;\;\; \psi=
\frac{m_2\xi_1-m_1\xi_2}{\lambda_{12}(m_1+m_2)} \, .\ee
In terms of new variables, the two body Lagrangian becomes a sum
of $L_{cm} $ and $L_{rel}$,
\be L_{cm}=\frac{1}{2}(m_1+m_2)\dot{\bR}^2+
\frac{1}{2(m_1+m_2)}\dot{\chi}^2\, , \ee
\be L_{rel}=
\frac{1}{2}\left(\mu+\frac{\lambda_{12}}{r}\right)\dot{\br}^2
+\frac{\lambda_{12}^2}{2}
\left(\mu+\frac{\lambda_{12}}{r}\right)^{-1}(\dot{\psi}+\bw(\br)\cdot\dot{\br})^2\,
. \ee

The requirement that relative moduli space of two space being
nonsingular is that the coupling constant $\lambda_{12}$ should be
positive and the relative coordinate $\psi$ has to have a period
of $4\pi$. Let us consider the range of the angle parameters. The
shift of $\xi_1$ by $4\pi t_1$ implies the identification
\be (\chi,\psi) = \left(\chi+ 4\pi t_1, \psi+\frac{4\pi
m_2t_1}{\lambda_{12}(m_1+m_2)}\right)\, , \ee
and the shift of $\xi_2$ by $-4\pi t_2$ implies the identification
\be (\chi,\psi) = \left(\chi- 4\pi t_2, \psi+\frac{4\pi
m_1t_2}{\lambda_{12}(m_1+m_2)} \right) \, .\ee
A combination of $\lambda_{12}/t_1$ steps of the first shift and
$\lambda_{12}/t_2$ steps of the second shift will lead to an
identification
\be (\chi,\psi)=(\chi,\psi+4\pi) \, .\ee
For this operation to be minimum so that the period of $\psi$ is
$4\pi$, rather than a smaller number, $\lambda_{12}/t_1$ and
$\lambda_{12}/t_2$ should be co-prime integers. There are several
consequences from this requirement. The quantization of charge
$q_i$ leads to the relative charge as
\be q_{rel}= \frac{\lambda_{12}}{2t_1t_2}(t_2n_1-t_1n_2)\, , \ee
with integer $n_1, n_2$. There are pair of integers such that
$q_{rel}=1/2$ as expected. The ratio of the periods $t_1/t_2$
should be a positive rational number. After scaling the
coordinates, we can make $t_1,t_2$ to be integers and
$\lambda_{12}$ to be the least common multiplet of $t_1,t_2$.

In short distance where $\bx_1$ and $\bx_2$ particles approach
each other, the corresponding metric becomes
\be {\cal G}_{\rm rel} = \frac{\lambda_{12}}{2}ds_{R^4}^2 \, ,\ee
where
\be ds_{R^4}^2 = \frac{1}{r}d\br^2+r (d\psi+\cos\theta d\phi)^2 \,
\ee
with the period of $\psi$ in $[0,4\pi]$ is the flat metric for
Euclidean four dimensional space.

Generalizing to the N distinct monopoles of $SU(N)$ gauge group,
the ratio of any pair of periods of monopoles in adjacent point in
the root diagram should be rational for the geometry to be
nonsingular when two monopoles are coming together. As all
monopoles are interacting each other at least indirectly, one can
scale space and time and so the periods $t_i$ are all integers,
without any common factor. In our case also one can argue that the
moduli space is smooth when $N-1$ distinct monopoles are coming
together by the argument similar to Ref.~\cite{Lee:1996kz}.
However, the space becomes singular when $N$ distinct monopoles
coming together as they form a generalization of the moduli space
of a single caloron of $SU(N)$ gauge group~\cite{Lee:1999xb}. Our
generalization of the monopole moduli space for $SU(3)$ would be
exactly the three parameter $(t_1,t_2,t_3)$ generalization of the
moduli space $N(1,1)$ of single $SU(3)$ instanton. In the massless
limit where two of monopole mass vanishes, the monopole moduli
space becomes $\CM_\bt$ after scaling of the coordinates.

\end{document}